\documentclass[aps,prd,a4paper,nofootinbib,
preprintnumbers,superscriptaddress,showpacs,showkeys,twocolumn]{revtex4-2}
 
\usepackage{amsmath}
\usepackage{amssymb}
\usepackage{bm}
\usepackage{graphicx}
\usepackage{xcolor}
\usepackage{slashed}
\usepackage{comment}
\usepackage[normalem]{ulem}

\usepackage{xcolor}
\usepackage[
colorlinks=true,
linkcolor=blue,
citecolor=blue,
urlcolor=blue,
breaklinks,
pdfstartview=FitH,
bookmarksopenlevel=0,
bookmarks=true,
pdftoolbar=true,
pdfauthor={{}}
pdftitle={Probing Criticality: Testing Lattice QCD Techniques with Effective Models},
pdfkeywords={FRG, fluid dynamics, effective field theories, QCD, Lattice QCD}
]{hyperref}

\usepackage[sort,compress]{cleveref}

\crefname{section}{Sec.\!}{Secs.\!}
\crefname{equation}{Eq.\!}{Eqs.\!}
\crefname{figure}{Fig.\!}{Figs.\!}
\crefname{table}{Tab.\!}{Tabs.\!}
\crefname{appendix}{App.\!}{Apps.\!}
\crefname{chapter}{Chapter}{Chapters}

\crefname{section}{Sec.\!}{Secs.\!}
\crefname{equation}{Eq.\!}{Eqs.\!}
\crefname{figure}{Fig.\!}{Figs.\!}
\crefname{table}{Tab.\!}{Tabs.\!}
\crefname{appendix}{App.\!}{Apps.\!}

\newcommand{\be}{\begin{equation}}
\newcommand{\ee}{\end{equation}}
\newcommand{\ba}{\begin{eqnarray}}
\newcommand{\ea}{\end{eqnarray}}

\begin{document}
\title{Assessing the Reconstruction of the Critical Line in the QCD Phase Diagram from Imaginary to Real Chemical Potential}

\author{Fabrizio Murgana}\email{fabrizio.murgana@dfa.unict.it}
\affiliation{Department of Physics and Astronomy "Ettore Majorana", University of Catania, Via Santa Sofia 64, I-95123 Catania, Italy}\affiliation{INFN-Sezione di Catania, Via Santa Sofia 64, I-95123 Catania, Italy}

\author{Marco Ruggieri}\email{marco.ruggieri@dfa.unict.it}
\affiliation{Department of Physics and Astronomy "Ettore Majorana", University of Catania, Via Santa Sofia 64, I-95123 Catania, Italy}\affiliation{INFN-Sezione di Catania, Via Santa Sofia 64, I-95123 Catania, Italy}


\begin{abstract}
We test a technique adopted in the lattice simulations framework, 
to reconstruct the chiral-phase boundary 
at real chemical potential, $\mu$,
via extrapolation from imaginary $\mu$. 
We use a low-energy effective model, the Quark-Meson model, both in the mean-field approximation and within the Functional Renormalization Group, the latter
in the Local Potential Approximation. 
The model provides   results both for real and imaginary values of $\mu$, thus a direct comparison can be performed between the prediction of the model for real values of $\mu$ and the ones obtained via extrapolation from the results at imaginary $\mu$. 
We compute an effective convergence radius for the reconstruction technique, $\mu_\mathrm{conv}$, and find $\mu_\mathrm{conv}\approx146$ MeV. This value sustains 
the validity of the reconstruction technique also for finite 
and moderate values of the chemical potential. On the other hand, 
within our model, $\mu_\mathrm{conv}$
is quite smaller then the value of 
$\mu$ where we find 
the actual critical endpoint. 
Near this point of the phase diagram,
we find a discrepancy between the actual phase boundary 
and the one pbtained via extrapolatin of $\approx150\%$. 
Therefore, our results show that the location of the critical 
endpoint obtained via reconstruction from imaginary $\mu$
should be considered with due caution.
\end{abstract}


\maketitle

\section{Introduction}
The study of Quantum Chromodynamics (QCD) under extreme conditions of temperature and baryon density remains one of the central challenges in contemporary physics. Of particular interest is the behavior of strongly interacting matter at finite temperature and chemical potential, as it governs the phase structure of QCD, including the possible existence of a critical endpoint (CEP) separating regions of crossover and first-order phase transition \cite{Fukushima_2011, Barducci:1989wi, Meyer:96, Herbst:2013ail, Murgana:2023pyx, Stephanov:1998dy,Wilczek:1992sf, Ruggieri:13}. This complex landscape is not only of theoretical significance but also directly relevant to experimental programs in heavy-ion collisions, where the search for signals of the QCD phase transition and the CEP constitutes a major objective \cite{Li:18, Roland:2014jsa, Proceedings:2007ctk, Xu:2014jsa} .

In this context, lattice QCD provides a first-principles framework for calculating the thermodynamics of QCD. At vanishing chemical potential, $\mu$, 
lattice simulations are firmly established and have produced highly accurate results for quantities such as the equation of state, fluctuations of conserved charges, and correlation functions \cite{Bazavov:12, Ratti:2004ra, Ratti:2005jh}. However, as the baryon chemical potential increases, lattice QCD calculations are hindered by the well-known sign problem, which severely restricts direct simulations in the high-density regime \cite{Pan:22}. This challenge has motivated the development of various indirect reconstruction techniques—such as Taylor expansions \cite{Gavai:2003mf,Bazavov:2017dus} , analytic continuation from imaginary chemical potential \cite{Alford:1998sd, Bellwied:2015rza}, and complex Langevin methods \cite{Parisi:1983mgm,Attanasio:2020spv} to extend lattice results to finite chemical potential.

Parallel to these efforts, a wide array of effective models has been developed to explore the thermodynamics of QCD in regions where lattice calculations become intractable. Models such as the Nambu–Jona-Lasinio (NJL) model \cite{Nambu:1961tp,Nambu:1961fr}, quark-meson models \cite{Koch:97ei, Weinberg:96}, Polyakov-loop extended chiral models \cite{Fukushima:2017csk} , and others attempt to capture essential features of QCD by incorporating key symmetries and interactions, often calibrated to reproduce lattice data at zero or low chemical potential. While these models lack the first-principles rigor of lattice QCD, they are invaluable tools for exploring qualitative trends, making predictions at high baryon density, and testing scenarios for the structure of the QCD phase diagram.

The main purpose of our work is to perform a comparative analysis between the reconstruction
of the critical boundary via continuation of the results
from imaginary to real $\mu$,
and the actual boundary computed at real $\mu$.
In particular, we want to check whether the continuation of
the critical line from imaginary to real $\mu$ gives,
within our model, a reasonable estimate of the 
location of the CEP at real $\mu$.
A similar study within the Dyson-Schwinger equation (DSE)
framework
can be found in~\cite{Bernhardt:2023ezo}.
By examining the phase diagram predicted by the model at both real and imaginary $\mu$, we assess the degree to which the reconstruction aligns with model predictions, and how this agreement evolves as one approaches the CEP. 
Such a comparison serves multiple purposes: 
in fact,
it validates the reconstruction technique in 
the small $\mu$ region, and identifies the limitations of the reconstruction as $\mu$ is increased and the CEP is approached.

Specifically, we exploit the Quark-Meson (QM) model \cite{Koch:97ei, Peskin:95, Gell-Mann:60, Weinberg:96} both in the MF and in the Hydrodynamical formulation of the Functional Renormalization Group (FRG) in the Local Potential Approximation (LPA) scheme. 
This technique has previously been applied to various effective models  and for calculations at real $\mu$ up to the critical endpoint (see, e.g., \cite{Murgana:2023xrq,Murgana:2023pyx, Jeong:2024rst, koenigstein:21, Braun:18, Stoll:21}). 
In the present work, 
we extend its application to imaginary  $\mu$. By performing calculations at imaginary $\mu$, then 
following a similar extrapolation procedure as employed in lattice QCD studies, we compare the extrapolated crossover transition line with the results obtained explicitly through our functional approach. 

Our results demonstrate a clear and systematic trend. 
At low $\mu$, the reconstructed critical line 
and the actual one exhibit a high degree of agreement, 
consistent with the expectation that both approaches are well-grounded in this regime. 
However, as $\mu$ increases and approaches 
the region associated with the CEP, the agreement begins to falter. The increasing discrepancy reflects the growing influence of critical fluctuations and non-perturbative dynamics that are challenging to capture by a simple extrapolation of the results
at imaginary $\mu$. 
Thus, our findings can serve as a benchmark for assessing the reliability of the extrapolation method commonly adopted in the lattice framework.

This article is organized as follows. In Section II, we describe the Quark Meson model as an effective low-energy description of chiral symmetry breaking in QCD. Section III provides an overview of the theoretical FRG setup used in this study. Section IV presents the comparative analysis of the results, with particular attention to the behavior of key observables as functions of chemical potential. Finally, in Section V, we summarize our findings and discuss their implications for future work on the QCD phase diagram and the search for the critical endpoint.

\section{Model setup}

The Quark-Meson (QM) model provides an effective framework to describe low-energy QCD, extending the linear sigma model by incorporating fermionic degrees of freedom \cite{Koch:97ei, Weinberg:96, Gell-Mann:60}. In the case of two quark flavors ($ N_f = 2 $), the fundamental fields of the QM model include an isospin triplet of pions, $ \vec{\pi} = (\pi_1, \pi_2, \pi_3) $, and a scalar isosinglet field, $ \sigma $. These mesonic fields couple to a massless isospin doublet fermion field, $ \psi $, which represents the up and down quarks. The Euclidean Lagrangian density for this model is given by  
\begin{align}
\mathcal{L}_{\text{E}}^{\text{QM}} &= \bar{\psi} \left[  \gamma_\mu \partial_\mu + h (\sigma + i \gamma_5 \vec{\tau} \cdot \vec{\pi} ) \right] \psi \\
&+ \frac{1}{2} (\partial_\mu \sigma)^2 + \frac{1}{2} (\partial_\mu \vec{\pi})^2 + U (\sigma^2 + \vec{\pi}^2) - c \sigma,
\end{align}  
where $ \gamma_\mu $ and $ \gamma_5 $ are the Euclidean Dirac matrices, $ h $ denotes the Yukawa coupling constant controlling the interaction between quarks and mesons, and $ \vec{\tau} = (\tau^1, \tau^2, \tau^3) $ represents the Pauli matrices in flavor space. The potential $ U(\sigma^2 + \vec{\pi}^2) $ is designed to maintain $ O(4) $ symmetry, as it depends solely on the invariant combination $ \sigma^2 + \vec{\pi}^2 $. However, if the potential acquires a minimum in a particular direction, spontaneous symmetry breaking occurs, reducing $ O(4) $ to $ O(3) $. This symmetry breaking allows one to choose a vacuum configuration in which  
\begin{equation}
\langle \vec{\pi} \rangle = 0, \quad \langle \sigma \rangle = f_\pi \neq 0,
\end{equation}  
where $ f_\pi = 0.093 $ GeV corresponds to the pion decay constant.  

Although the Lagrangian does not include an explicit mass term for the quarks, spontaneous symmetry breaking generates a dynamical mass for them, given by  
\begin{equation}
M = h \langle \sigma \rangle.
\end{equation}  
Moreover, the term $ -c\sigma $ explicitly breaks $ O(4) $ symmetry, effectively mimicking the presence of a small but finite current quark mass. Due to this explicit symmetry breaking, the pions acquire a finite mass, behaving as pseudo-Goldstone bosons. Their mass is given by  
\begin{equation}
M_\pi^2 = \frac{c}{f_\pi}.
\end{equation}

\section{Functional Renormalization Group  Framework}

In this section, we provide a concise overview of the Functional Renormalization Group (FRG) approach, following the formulation introduced by Wetterich and collaborators~\cite{Ellwanger:1993mw, Wetterich:1992yh,Morris:94,Reuter:1993kw}. For a more detailed discussion, the reader is referred to Refs. \cite{Pawlowski:2005xe, koenigstein:21,Kopietz:2010zz,Dupuis:2020fhh,Berges:02}.  

The central quantity in this framework is the effective action, $ \Gamma[\Phi] $, which serves as the generator of one-particle irreducible (1PI) vertex functions. To facilitate its computation, an intermediate quantity known as the effective average action, $ \bar{\Gamma}_k[\Phi] $, is introduced. This function depends on a momentum-like parameter $ k $, which acts as a coarse-graining scale. The evolution of $ \bar{\Gamma}_k[\Phi] $ is described by the FRG flow equation, governing its transition from the ultraviolet (UV) regime ($ k \to \infty $) to the infrared (IR) limit ($ k \to 0 $). The effective average action smoothly interpolates between the bare classical action, $ S_{\text{bare}}[\Phi] $, at high energies and the full quantum effective action, $ \Gamma[\Phi] $, in the low-energy limit:

\begin{equation}
\bar{\Gamma}_{k \to \infty}[\Phi] = S_{\text{bare}}[\Phi], \quad \bar{\Gamma}_{k \to 0}[\Phi] = \Gamma[\Phi].
\end{equation}

In practical applications, it is not always feasible to take $ k \to \infty $. Instead, a finite UV cutoff $ \Lambda $ is introduced as the initial scale for the FRG flow. This cutoff is chosen to be significantly larger than any other physical scale in the theory to ensure that the bare classical action accurately describes the system at this energy scale. However, since using a finite cutoff introduces an approximation, it is crucial to verify that physical results remain independent of $ \Lambda $, which can be achieved by maintaining renormalization group (RG) consistency (see, for example \cite{Braun:18}).

The evolution of $ \bar{\Gamma}_k[\Phi] $ as fluctuations with decreasing momenta are integrated out is described by the Wetterich equation, also known as the Exact Renormalization Group (ERG) equation \cite{WETTERICH:91,Wetterich:1992yh}
\begin{equation}
\partial_k \bar{\Gamma}_k[\Phi] = \text{Tr} \left[ \frac{1}{2} \partial_k R_k \left( \bar{\Gamma}^{(2)}_k[\Phi] + R_k \right)^{-1} \right].
\end{equation}
Here, the trace represents a summation over all internal degrees of freedom as well as an integration over momenta. This equation exhibits a one-loop structure since it contains the full propagator, which is given by  
\begin{equation}
G_k[\Phi] = \left( \bar{\Gamma}^{(2)}_k[\Phi] + R_k \right)^{-1},
\end{equation}
where $ R_k $ is the regulator function.  The regulator $ R_k $ is chosen to ensure a smooth transition between the bare action and the full effective action, imposing that $ R_k $ should act as a mass term for low-momentum modes ($ p < k $), regularizing the propagator in the IR, and 
     vanish in the limit $ k \to 0$ so that the full quantum effective action is recovered.
    Furthermore $ R_k $ should diverge as $ k \to \infty $ to ensure that the functional integral defining $ \bar{\Gamma}_k[\Phi] $ is dominated by the bare action at high energies.

The precise form of $ R_k $ is problem-dependent and can be optimized according to various criteria \cite{PAWLOWSKI:07,Litim:2002cf,Litim:2001up,Canet:2002gs}.  
Although the Wetterich equation appears compact, it is a functional integro-differential equation and cannot be solved exactly in general. To make progress, approximations must be introduced. Two widely used approaches are the vertex expansion \cite{Morris:1993qb,Bergerhoff:98,Morris:94} and the derivative expansion \cite{Berges:95, Berges:02}:  
In the vertex expansion, $ \bar{\Gamma}_k[\Phi] $ is expanded in powers of the fields, with the expansion coefficients given by the $ n $-point vertex functions:
    \begin{equation}
    \bar{\Gamma}_k^{(n)}(x_1, \dots, x_n) = \frac{\delta^n \bar{\Gamma}_k[\Phi]}{\delta \Phi(x_1) \cdots \delta \Phi(x_n)} \Bigg|_{\Phi=\Phi_0}.
    \end{equation}
    In the derivative expansion, $ \bar{\Gamma}_k[\Phi] $ is expressed as a series of composite operators involving derivatives of the fields. These terms must respect the symmetries of the theory.

Both expansions lead to an infinite hierarchy of coupled integro-differential equations, which must be truncated at a given order. In this work, we focus on the derivative expansion, as it remains valid even near phase transitions, where non-analyticities may arise in the effective action. In contrast, the vertex expansion assumes regularity of $ \bar{\Gamma}_k[\Phi] $, which can break down at critical points due to singularities or discontinuities in the FRG flow (see, e.g.,  \cite{Grossi:19,Grossi:2021ksl,koenigstein:21, Stoll:21, Murgana:2023pyx,Murgana:2023xrq}).  

\begin{widetext}

Retaining the full advantage of the properties of the derivative expansion,
we adopt the lowest-order truncation, commonly referred to as the Local Potential Approximation (LPA) \cite{Berges:02, PAWLOWSKI:07, Delamotte:2007pf}. In this work, we apply the LPA ansatz to the Quark-Meson (QM) model at finite temperature and nonzero quark chemical potential, which leads to the effective action
\begin{align}
\Gamma_k[\bar{\psi}, \psi, \phi] &=
\int_0^{\beta} dx_4 \int d^3x~\bar{\psi}
\left[
 \gamma^\mu \partial_\mu + h(\sigma + i \gamma_5 \vec{\tau} \cdot \vec{\pi})  -\mu \gamma^0 \right] \psi\nonumber\\
&
+ 
\int_0^{\beta} dx_4 \int d^3x~
\left[\frac{1}{2} (\partial_\mu \sigma)^2 + \frac{1}{2} (\partial_\mu \vec{\pi})^2 + U_k(\sigma^2 + \vec{\pi}^2) - c\sigma\right].
\end{align}
To regulate the flow equations, we employ the three-dimensional Litim regulator \cite{Litim:01, Litim:02}, applied separately to bosons and fermions, namely
\begin{align}
R_{k,B}(p) &= (k^2 - p^2) \Theta(k^2 - p^2), \\
R_{k,F}(p) &= i \not{p} \left( \frac{k^2}{p^2} - 1 \right) \Theta(k^2 - p^2).
\end{align}
Here, the subscripts $B$ and $F$ denote boson and fermion 
regulators respectively.
Using this regulator, the flow equation for the effective potential
reads
\begin{equation}
\partial_t U_k(\sigma) = - \frac{k^5}{12\pi^2} 
\left\{
\left[
\frac{1}{E_{k,\sigma}} \coth\left(\frac{E_{k,\sigma}}{2T}\right) 
+ \frac{3}{E_{k,\pi}} \coth\left(\frac{E_{k,\pi}}{2T}\right)
\right]
+ \frac{4N_c}{E_\psi} 
\left[
\tanh\left(\frac{E_\psi - \mu}{2T}\right) + \tanh\left(\frac{E_\psi + \mu}{2T}\right)
\right]\right\}.
\label{eq:flow}
\end{equation}
Here, $ t = -\ln(k/\Lambda) $, 
while the energy terms are defined as
\begin{align}
E_{k,\sigma} &= \sqrt{k^2 + \partial^2_\sigma U_k(\sigma)}, \\
E_{k,\pi} &= \sqrt{k^2 + \frac{\partial_\sigma U_k(\sigma)}{\sigma}}, \\
E_\psi &= \sqrt{p^2 + M^2},
\end{align}
and the constituent quark mass is given by $ M = h\langle \sigma \rangle $.

We now follow the prescription given in previous works, see, e.g., \cite{Zorbach:2024rre, koenigstein:21,Koenigstein:2021llr,Murgana:2023xrq,Grossi:19}, and define:
\begin{align}
u_k(\sigma) &= \partial_\sigma U_k(\sigma), \\
u'_k(\sigma) &= \partial_\sigma u_k(\sigma).
\end{align}
Taking the derivative of the flow equation with respect to $ \sigma $, we obtain an advection-diffusion equation with a source term \cite{Koenigstein:2021llr, Murgana:2023pyx, Grossi:2021ksl}:
\begin{equation}
\label{eq:flowder}
\partial_t u_k(\sigma) + \partial_\sigma f_k(\sigma, u_k(\sigma)) = \partial_\sigma g_k(u'_k(\sigma)) + N_c \partial_\sigma S_k(\sigma),
\end{equation}
where the different contributions are defined as follows.
Firstly,
the advection flux is
\begin{equation}
    f_k(\sigma, u_k(\sigma)) = \frac{k^5}{4\pi^2} \frac{1}{E_{k,\pi}} \coth\left(\frac{E_{k,\pi}}{2T}\right).
\end{equation}
The next contribution is the diffusion flux,
\begin{equation}
    g_k(u'_k(\sigma)) = -\frac{k^5}{12\pi^2} \frac{1}{E_{k,\sigma}} \coth\left(\frac{E_{k,\sigma}}{2T}\right).
\end{equation}
Finally, the source term is
\begin{equation}
    N_c S_k(\sigma) = \frac{N_c k^5}{3\pi^2} \frac{1}{E_\psi}
    \left[
    \tanh\left(\frac{E_\psi - \mu}{2T}\right) + \tanh\left(\frac{E_\psi + \mu}{2T}\right)
    \right].
\end{equation}
\end{widetext}

Within this framework, the function $ u_k(\sigma) $ 
plays the role of a conserved quantity, as it obeys a generalized conservation law (see e.g. \cite{Kurganov:00, Thomas:99,BLADE:17,Tadmor:98} for a general review on the topic, and \cite{Koenigstein:2021syz, Koenigstein:2021llr,Koenigstein:2021rxj,Ihssen:2023qaq} for FRG applications). 
Each term in the equation originates from a different physical process.
The advection flux arises from the pion sector. The presence of the factor of 3 in Eq. (5) confirms this contribution. Since the pion mass term $ u_k(\sigma)/\sigma $ vanishes at the minimum of the effective potential, pions behave as Goldstone bosons. Furthermore, the characteristic speed $ \partial_u f_k(\sigma, u_k(\sigma)) $ is positive for $ \sigma < 0 $ and negative for $ \sigma > 0 $, meaning that $ u_k(\sigma) $ and the minimum of the potential are advected toward lower values of $ \sigma $.
The diffusion term is induced by the radial sigma mode and depends on the curvature $ u'_k(\sigma) $. Diffusion acts to smooth out gradients in $ u_k(\sigma) $, thereby suppressing sharp features and discontinuities.
Finally, the source term arises from the fermionic loop, leading to a time- and $ \sigma $-dependent modification of the flow equation. Unlike the other contributions, this term is independent of $ u_k(\sigma) $ itself, reinforcing its role as an external source.

As a final comment,
we point out that the standard mean-field (MF) approximation can be formally recovered from \cref{eq:flow} and \cref{eq:flowder} by properly rescaling the flow equation and taking the $N_c\to \infty$ limit.

\section{Results}

In this section, 
we present the results obtained both in the mean field 
and in the FRG formulation. 
The pseudo-critical phase boundary within the model can be easily obtained by locating the pseudo-critical temperature, for a fixed either real or imaginary chemical potential, via the peak of 
\begin{equation}
    \chi_\sigma=-\frac{\partial \sigma}{\partial T },
    \label{eq:chi_sigma_111}
\end{equation}
where $\sigma$ indicates the expectation value of the sigma field obtained as the position of the minimum of the effective potential. In correspondence of a real second-order phase transition $\chi_\sigma$ diverges, thus we identify the pseudocritical temperature as the location of the peak of $\chi_\sigma$ as a function of temperature.
The resulting phase boundaries, obtained both in the MF and FRG formulations of the QM model are depicted in  \cref{fig:tc}.

\begin{figure}[t!]
    \centering
    \includegraphics[width=1\linewidth]{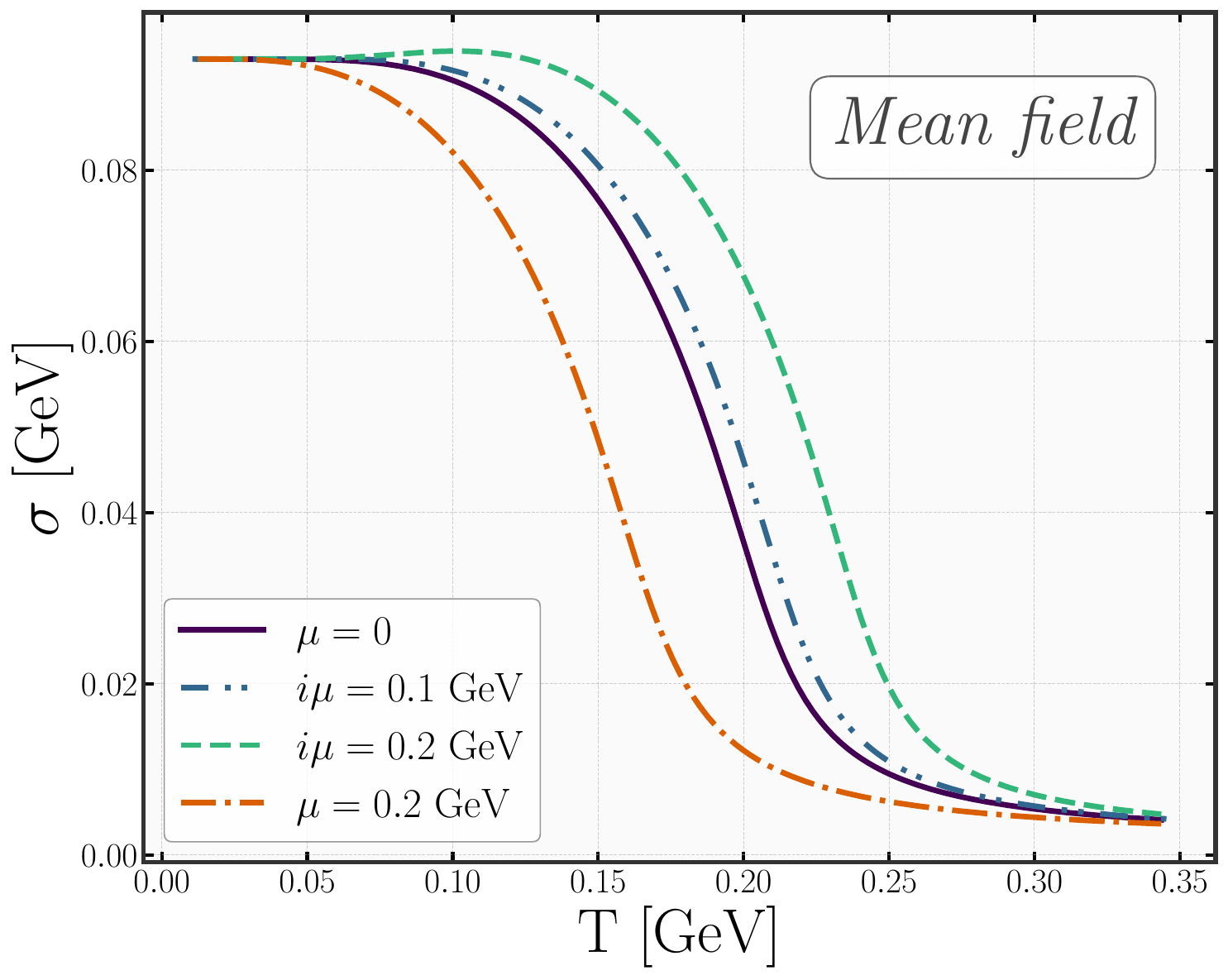}
    \includegraphics[width=1\linewidth]{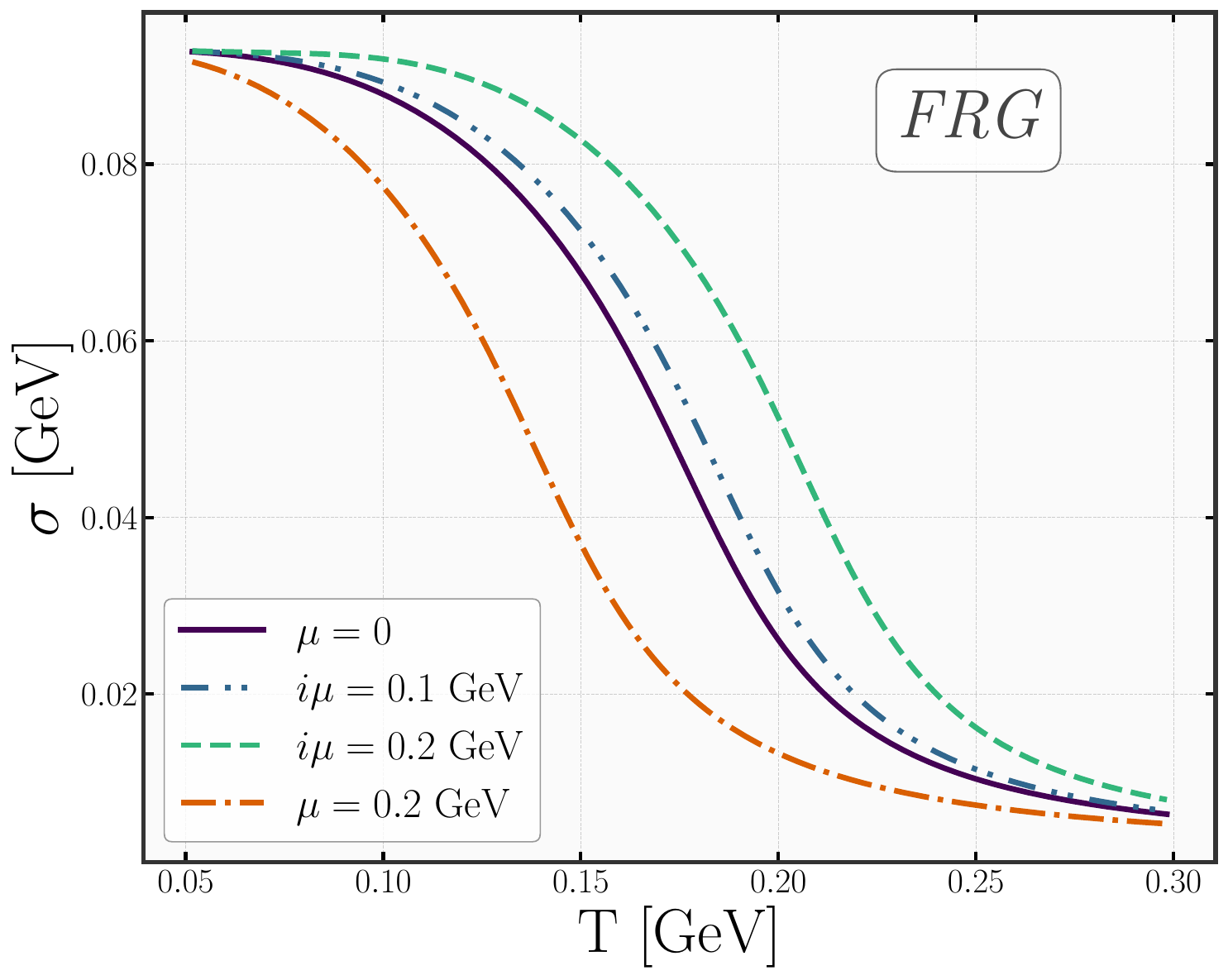}
    \caption{Expectation value of the $\sigma$ field as a function of temperature $T$ for different imaginary and real values of the chemical potential $\mu$ in the MF case (upper panel) and in the FRG case (lower panel). }
    \label{fig:min}
\end{figure}

In Fig.~\ref{fig:min} we plot $\sigma$ versus temperature
for $\mu=0$ (solid line), $\mu=0.2$ GeV
(dot-dashed line), then $i\mu=0.1, 0.2$ GeV 
(dot-dot-dashed and dashed line respectively).
Upper panel corresponds to the results obtained within the
MF approximation, while in the lower panel we show the
results obtained within the FRG scheme.
As expected, for a given value of the chemical potential,
increasing temperature results in the lowering of $\sigma$.
We can identify a temperature range in which the lowering of
$\sigma$ is more pronounced. This allows us to 
define the pseudo-critical temperature by looking at the
maximum of $\chi_\sigma$ defined in Eq.~\eqref{eq:chi_sigma_111}.
Increasing $\mu$ along the real axis, the pseudo-critical 
temperature decreases. On the other hand, increasing $\mu$
along the imaginary axis results in the increase of
the critical temperature. The behavior 
of $\sigma$ for the latter case that we have found
is in agreement with previous results, see for example \cite{Morita:2011jva}.

\begin{figure}[t]
    \centering
    \includegraphics[width=1\linewidth]{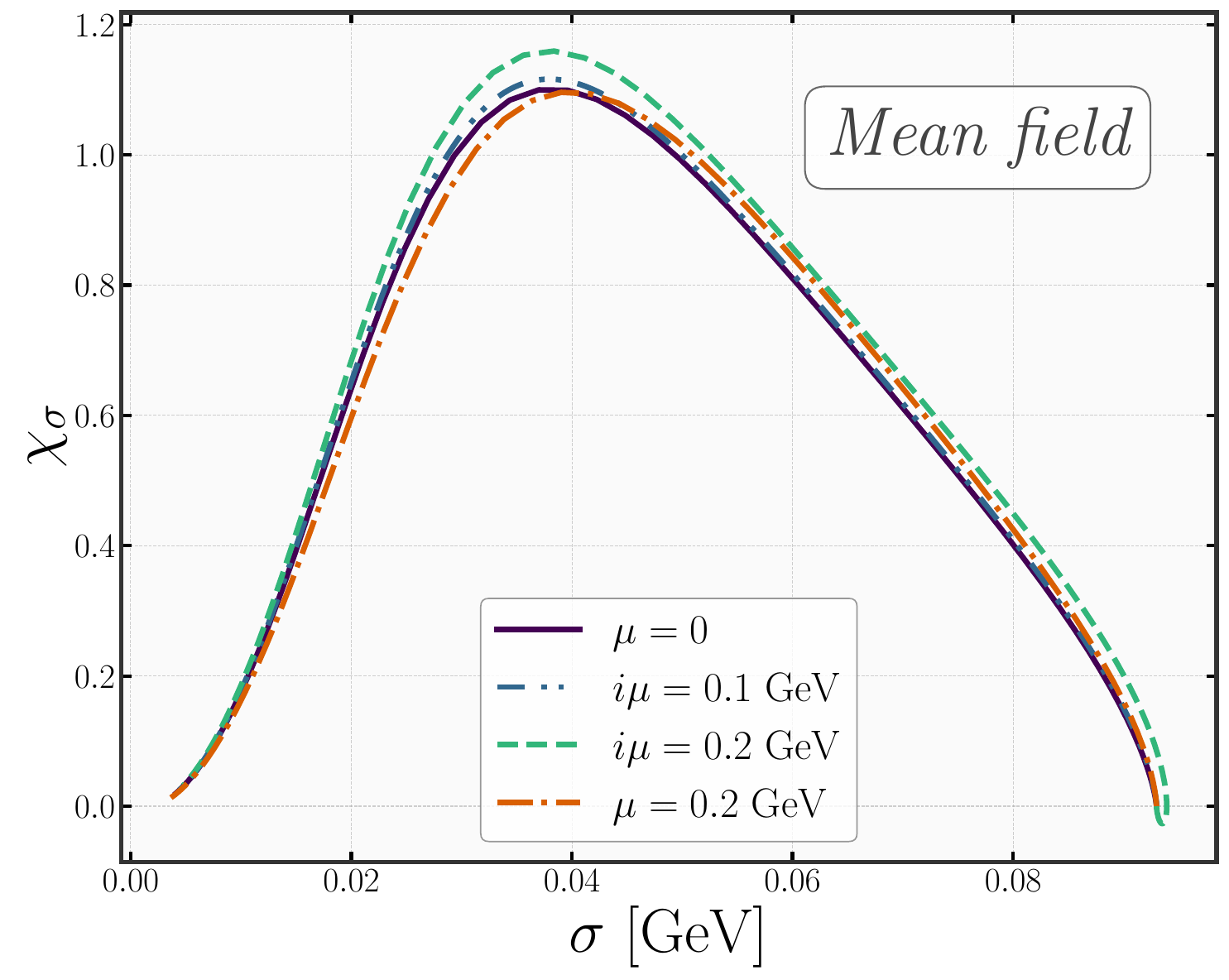}
    \includegraphics[width=1\linewidth]{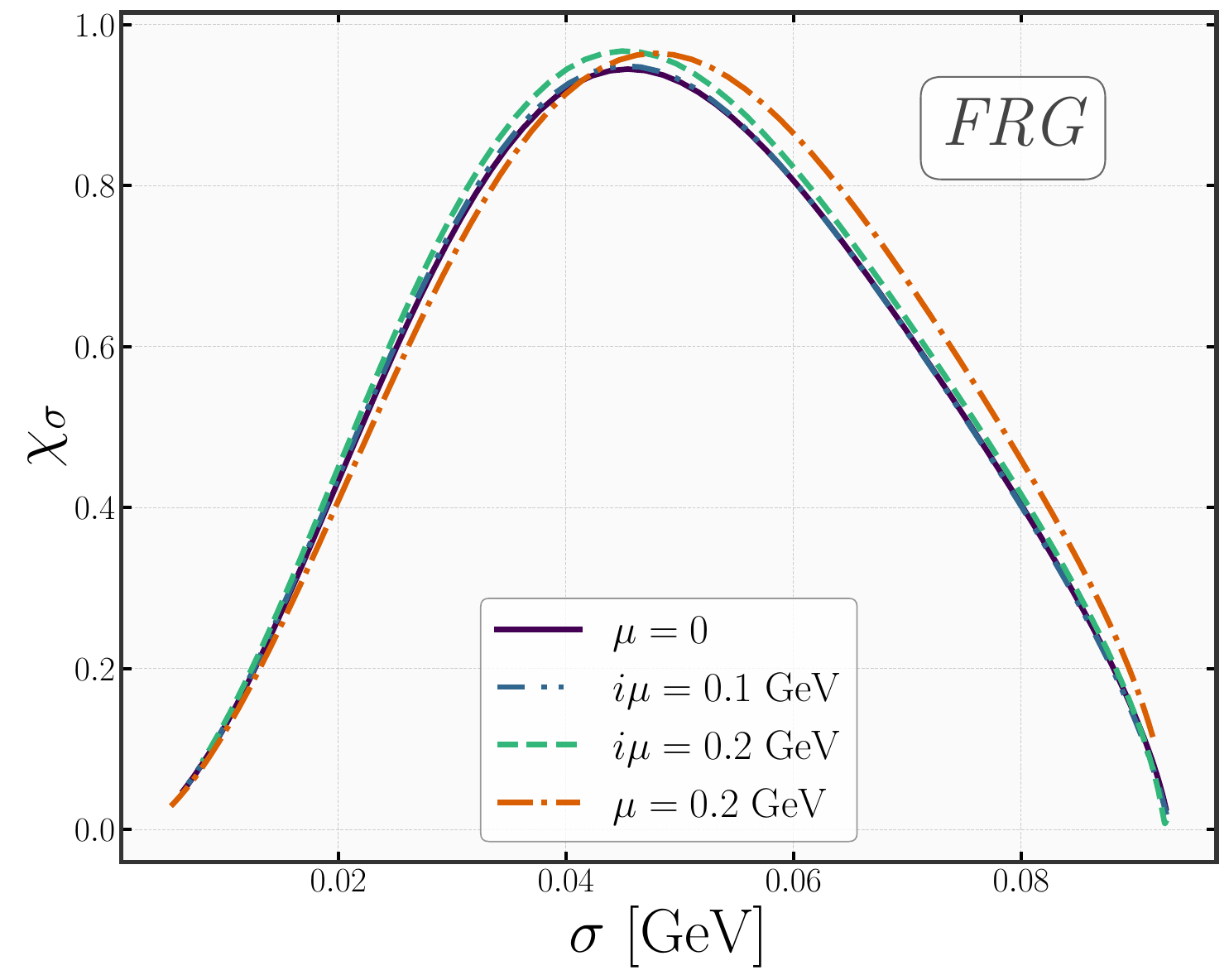}
    \caption{$\chi_\sigma$  as a function of the expectation value of the $\sigma$ field for different imaginary and real values of the chemical potential $\mu$ in the MF case (upper panel) and in the FRG case (lower panel). }
    \label{fig:chi}
\end{figure}

Our analysis on the pseudo-critical temperature at fixed  
chemical potential
is completed with the results shown in Fig.~\ref{fig:chi},
where we plot $\chi_\sigma$ versus $\sigma$.
Upper and lower panels correspond to results
obtained within MF approximation and FRG scheme
respectively.
In order to compile the data in Fig.~\ref{fig:chi},
for a given value of $\mu$,
$\sigma$ is computed at a temperature via minimization
of the effective potential, then $\chi_\sigma$ is computed
at the same temperature.
This figure is analogous to Fig. 2 of \cite{Bernhardt:2023ezo}.
The behavior of $\chi_\sigma$ 
versus $\sigma$ allows us to identify the temperature at which
$\chi_\sigma$ has its maximum.

\begin{figure}[t]
    \centering
    \includegraphics[width=1\linewidth]{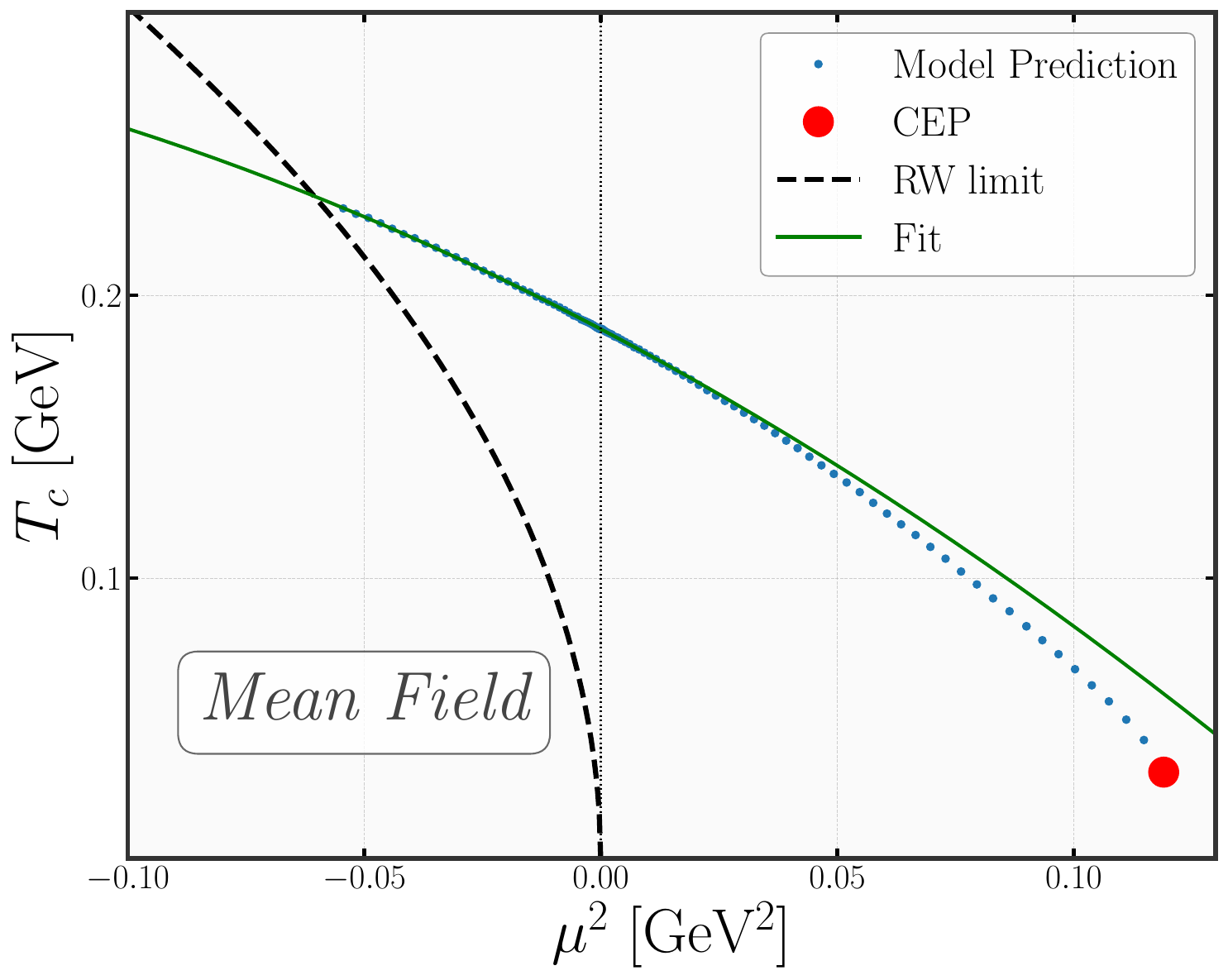}
     \includegraphics[width=1\linewidth]{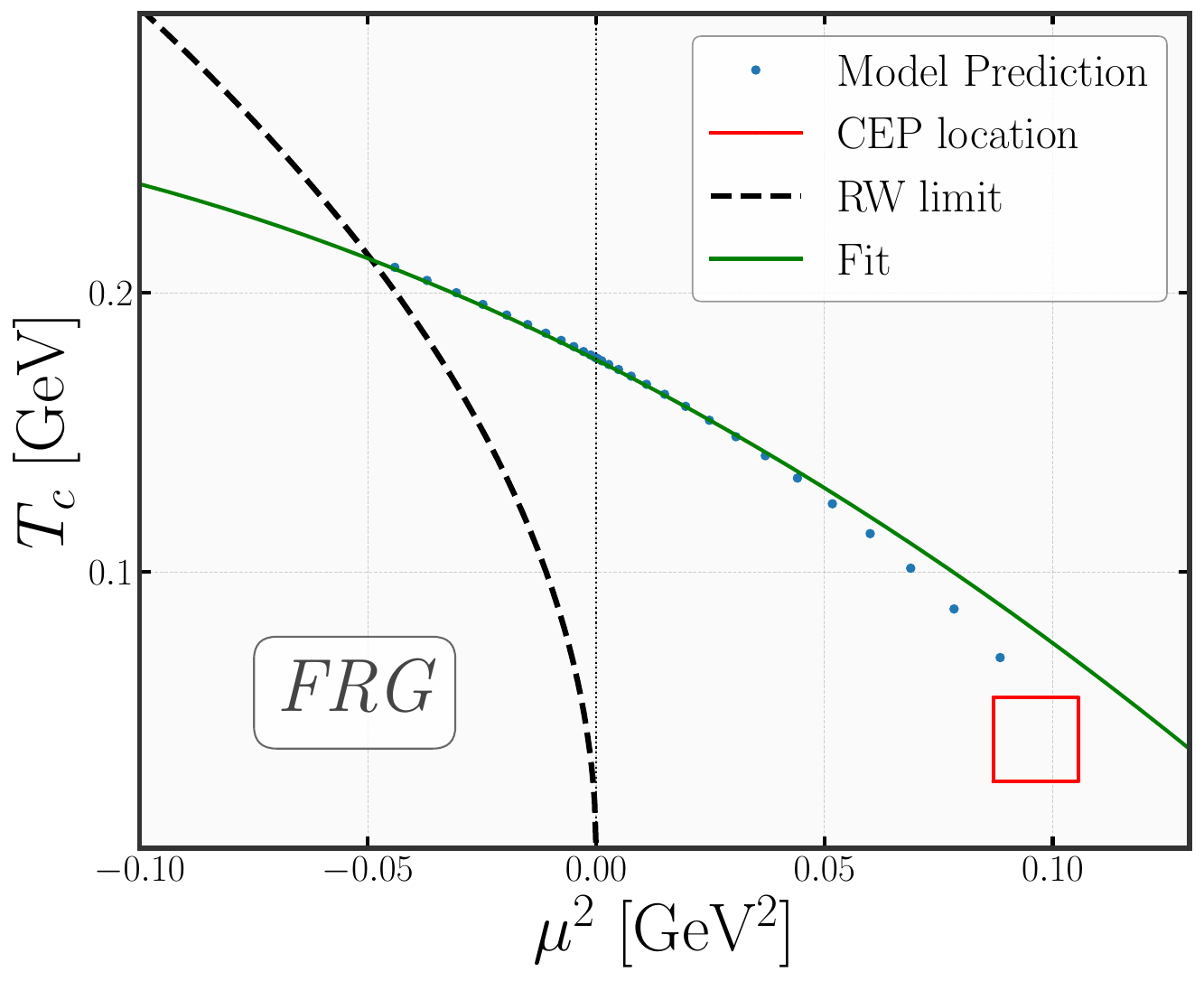}
    \caption{Phase boundary of the QM model in the mean field approximation (upper panel) and in the FRG case(lower panel), for both real and imaginary chemical potential $\mu$. The blue dots indicate the model prediction while the continuous green line indicates the reconstruction obtained via the techniques adopted in the lQCD framework based on imaginary chemical potential results. The red dot in the upper panel indicates the location of the CEP predicted by the model, while in the lower panel the red box includes also the uncertainties on its location due to the backbending phenomenon. }
    \label{fig:tc}
\end{figure}

\begin{figure}[t]
    \centering
    \includegraphics[width=1\linewidth]{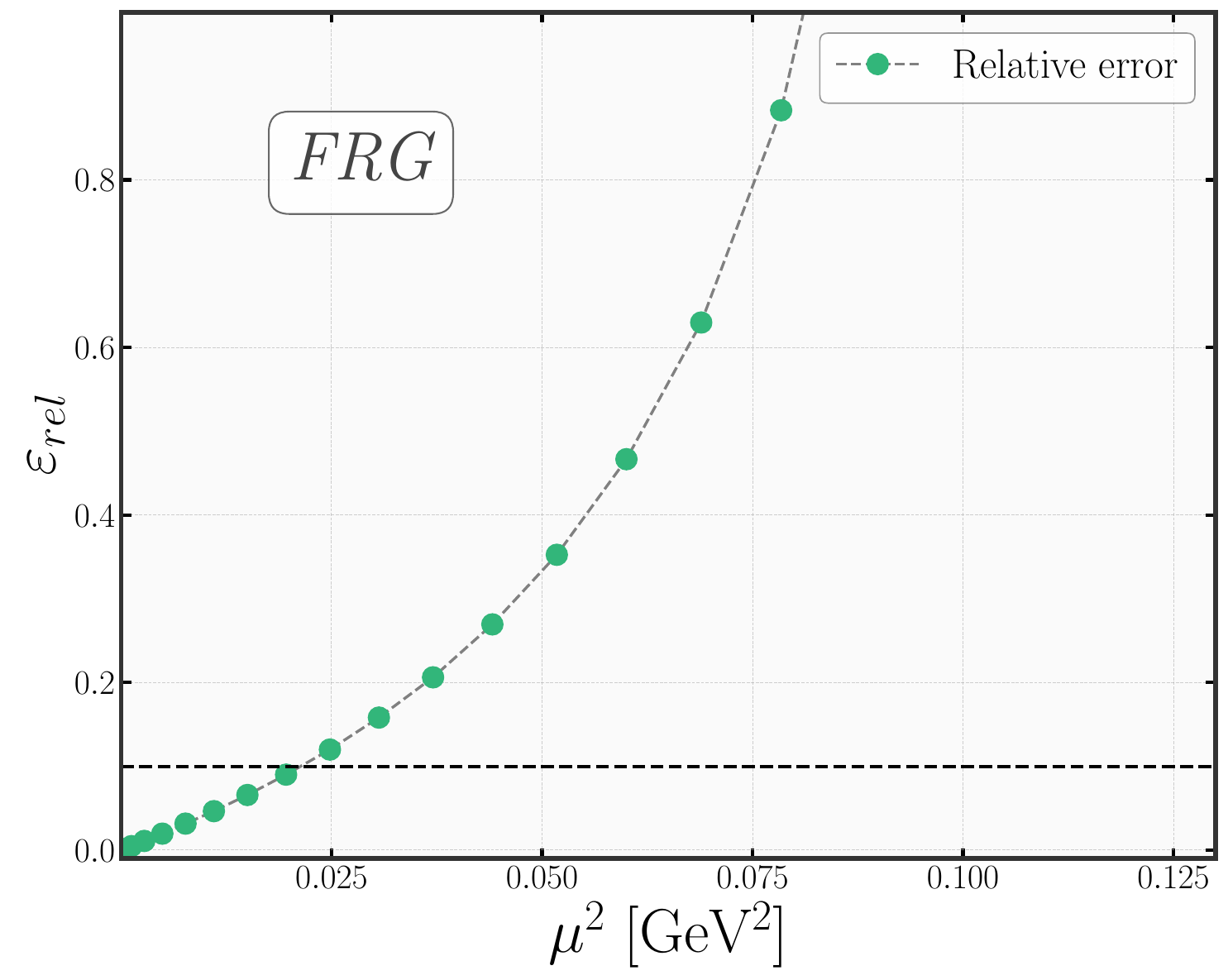}
 \caption{Relative error $\varepsilon_{rel}$ (see \cref{eq:err_rel}) of the extrapolated critical temperature in the FRG case. 
 The black horizontal line indicates the threshold of 0.1 in the relative error that we used to estimate the effective convergence radius of the reconstruction technique.  
 For the mean-field case the results are quantitatively similar.}
    \label{fig:err_rel}
\end{figure}

Next we focus on
the pseudo-critical line.
Our main purpose is to 
compare the results obtained through extrapolation 
from imaginary to 
real chemical potential, with those directly
obtained
by the model computations.

In \cref{fig:tc}, we display the pseudocritical temperature derived  at both imaginary and real chemical potentials. 
The results have been obtained by looking at the location of the
maxima of $\chi_\sigma$ as explained above.
The blue dots denote results from model computations, the dashed black line corresponds to the RW limit. 
The solid line corresponds to
a numerical fit of $T_c(\mu)$
with 
the standard functional form~\cite{Borsanyi:2020fev,Bellwied:2015rza}:
\begin{equation}\label{eq:fit}
\frac{T_c(\mu)}{T_c} = 1 - \kappa_2 \left(\frac{\mu}{T_c}\right)^2 - \kappa_4 \left(\frac{\mu}{T_c}\right)^4,
\end{equation}
where $T_c = T_c(\mu = 0)$, 
$\kappa_2$ and $\kappa_4$ quantify the curvature and
the kurtosis of the phase boundary.  
We remark that
in the literature, these coefficients are usually computed using the baryon chemical potential $\mu_b=3\mu$ instead $\mu$. This implies that they have to be adjusted in case one
is interested in a quantitative comparison with other
studies. For the purpose of the present study however,
this difference is irrelevant.  
For completeness,
we collect the parameters $\kappa_2$ and
$\kappa_4$
in \cref{tab:values} for MF and FRG calculations.

\begin{table}[t]
    \centering
    \begin{tabular}{|c|c|c|c|}
    \hline
        & $T_c$ (GeV)& $\kappa_2$& $\kappa_4$\\
        \hline 
        MF & $0.188$  & $0.0311\pm 0.0001$&
 $0.0021\pm0.0001$ \\
        FRG & $0.176$ & $
 0.0255\pm0.0005$ & $
 0255\pm0.0005$\\
        \hline
    \end{tabular}
    \caption{Values of the fit parameters for the MF and FRG cases. }
    \label{tab:values}
\end{table}

Since in the model calculations we have access
to the critical temperature both for real and imaginary
chemical potential, we can compare the actual results obtained
for $T_c$ at real $\mu$ with those obtained 
reconstructing $T_c$ by using Eq.~\eqref{eq:fit},
with $\kappa_2$ and $\kappa_4$ computed from the results
obtained at imaginary chemical potential and then continued 
to real $\mu$.

Notably, we observe both in the FRG and in the MF case, that the agreement between the two methods is significantly better in the regime of low chemical potential, particularly in the vicinity of vanishing chemical potential. In this region, the phase boundary reconstructed from the data at imaginary $\mu$ closely follow the model prediction for real chemical potentials, with discrepancies remaining well within acceptable uncertainties. 

However, as the chemical potential increases and approaches the conjectured location of the critical endpoint (CEP), the 
agreement between the reconstruction 
based on Eq.~\eqref{eq:fit}
and the model begins to deteriorate. 
This happens both in the MF and in the FRG cases.  We point out that, while in the MF case we have a precise location of the CEP, the FRG prediction accuracy is limited by the well-known  backbending \cite{Tripolt:18}: for high $\mu$ and low $T$, the phase 
diagram computed within FRG in the LPA
exhibits a positive slope, leading to a negative entropy density region. Within this region, the $\sigma$ expectation value is not monotonous and $\chi_\sigma$ exhibits both a minimum and a maximum, without reaching the divergence condition required for a pure $2^{nd}$-order phase transition. Thus, we can only limit the region in which the CEP is located considering the values of $\mu$ and $T$ for which we are clearly in a crossover and a $1^{st}$-order phase transition regime. 

The difference 
between the extrapolated result and the actual calculation
of the critical line
becomes more pronounced the closer we move toward the critical region. The progressive mismatch is expected,
as 
the critical line~\eqref{eq:fit}
is based on a reconstruction which is well defined 
in a neighborhood of $\mu=0$. 

In order to estimate the range of validity of the reconstruction technique from the data at imaginary chemical potential, 
we compute the relative difference between the 
actual values of the critical temperature predicted by the model, $T_c$, and the ones obtained via the 
reconstruction based on Eq.~\eqref{eq:fit},
$T_c^{(fit)}$, that is 
\begin{equation}\label{eq:err_rel}
    \varepsilon_{rel}=\frac{|T_c-T_c^{(fit)}|}{T_c}.
\end{equation}
In \cref{fig:err_rel}  we show 
$\varepsilon_{rel}$ as a function of $\mu^2$. 
As expected, the relative error increases with $\mu^2$.

We can use the results in \cref{fig:err_rel}
to extract an effective convergence radius 
in the $\mu^2$ axis
for the extrapolation technique. 
To this end, we have to fix a threshold value for 
$\varepsilon_{rel}$, above which we state that
the extrapolation~\eqref{eq:fit} is not convergent.
For the sake of concreteness,
we fix the threshold value for $\varepsilon_{rel}$ to $0.10$. 
This leads to the following effective convergence radii, $\mu_{conv}$,
respectively for the MF and the FRG cases:
\begin{equation}
    \mu_\mathrm{conv}=0.1463 \mbox{  GeV, } \qquad \mathrm{MF},
\end{equation}
and
\begin{equation}
    \mu_\mathrm{conv}= 0.1461 \mbox{  GeV, }\qquad \mathrm{FRG}.
\end{equation}
These results show that the effective
convergence radius is the same for both FRG and MF, within numerical
accuracy. However,  in both cases $\mu_{conv}$ is significantly smaller than the value corresponding to the location of the critical endpoint predicted by the model, which is more than twice $\mu_{conv}$,  and the relative error is of order $\varepsilon_{rel}\approx 1.5$ in the proximity of the CEP. 
This exercise suggests  
that the
extrapolation from imaginary to real (and large) 
baryon chemical potential,
in order to estimate the location of the critical endpoint,
should be taken with a grain of salt.

\section{Conclusions}

In this study, we have
adopted the hydrodynamic formulation of the FRG applied to the QM model, as an effective chiral model of QCD. We  have performed
a comparative analysis between 
the extrapolated critical line
from imaginary to real quark chemical potential, $\mu$,
and the one actually
computed at finite $\mu$.
Our objective was
to test the standard techniques utilized to reconstruct the phase boundary at real chemical potential starting from data at imaginary chemical potential. To this end, the quark-meson model has been 
used, both in the mean field 
and in the FRG approach. For the latter, we have implemented
a local-potential approximation.
The model provides results for the phase boundary both at real and imaginary $\mu$; hence, 
we have access to a direct comparison between the real $\mu$ results, predicted by the model, and the ones obtained via extrapolation from imaginary $\mu$. 

The results of our analysis show that at low chemical potential, particularly in the vicinity of \(\mu = 0\), the agreement between the reconstruction from imaginary $\mu$ 
and the model predictions is excellent.  

On the other hand, 
as we move toward higher values of $\mu$, 
and particularly as we approach the  CEP, a progressive deviation between the two approaches becomes evident. 
The phase boundary reconstructed by the 
extrapolation from imaginary $\mu$ data
diverges from the model predictions as $\mu$ is increased, 
indicating difficulty of the extrapolation
to capture the non-analytic behavior associated with criticality. 

The trend observed in the comparative analysis underscores both the strengths and the limitations of each approach. The good agreement at low chemical potentials reinforces the validity of the reconstruction method in the crossover regime, while the growing discrepancy near the CEP highlights the need for improved techniques to handle critical dynamics, whether through enhanced lattice methods, better model parameterizations, or hybrid approaches.
Importantly, this trend is not merely a technical shortcoming, since it reflects the fundamental challenges involved in describing critical phenomena in QCD. Near the CEP, long-range correlations and non-perturbative effects dominate, and any method not explicitly constructed to handle such dynamics is likely to show signs of breakdown.

It is our opinion that the present work 
reinforces the usefulness of comparative studies in understanding the strengths and limitations of different theoretical tools used to explore the QCD phase diagram. 
In particular,
while the agreement at low chemical potential supports the 
methodology of extrapolation from imaginary $\mu$, 
the discrepancy 
between the extrapolated critical line and the actual one
near the CEP highlights 
the need for improved techniques, 
based on first-principle calculations, for finite-density QCD.

Further studies can be carried out investigating  the deconfinement 
QCD phase transition, exploiting the Polyakov-loop enhanced version of the QM model, in which a more precise reconstruction of the imaginary chemical potential dynamic and the RW transition are possible. We intend to asses this problem in a future work.

\subsection*{Acknowledgments}
M. R. acknowledges Bruno Barbieri, Lautaro Martinez, Francesco Acerbi 
and John Petrucci for inspiration.
This work has been partly funded by 
the European Union – Next Generation EU through the research grants number P2022Z4P4B “SOPHYA - Sustainable Optimised PHYsics Algorithms: fundamental physics to build an advanced society”, 
under the program PRIN 2022 PNRR of the Italian Ministero dell’Università e Ricerca (MUR).

\bibliography{biblio1}

\end{document}